\documentclass{article}
\usepackage[utf8]{inputenc}
\usepackage[margin=1in]{geometry}
\usepackage{amsmath}
\usepackage{physics}
\usepackage{graphicx}
\usepackage{amssymb}
\usepackage{color}
\usepackage{appendix}
\usepackage{authblk}
\usepackage{outlines}
\usepackage{enumitem}
\setenumerate[1]{label=\arabic*}
\setenumerate[2]{label*=.\arabic*}
\setenumerate[3]{label*=.\arabic*}
\setenumerate[4]{label*=.\arabic*}

\newcommand{\SFF}{\text{SFF}}
\newcommand{\bbeta}{{\mathfrak b}}

\newcommand{\LSFF}{\text{LSFF}}

\title{The Loschmidt Spectral Form Factor}
\author[1]{Michael Winer}
\author[2]{Brian Swingle}
\affil[1]{Joint Quantum Institute, Department of Physics, University
of Maryland, College Park, Maryland 20742, USA}
\affil[2]{Department of Physics, Brandeis University, Waltham, Massachusetts 02453, USA}

\date{October 2021}

\begin{document}

\maketitle
\begin{abstract}
    The Spectral Form Factor (SFF) measures the fluctuations in the density of states of a Hamiltonian. We consider a generalization of the SFF called the Loschmidt Spectral Form Factor, $\tr[e^{iH_1T}]\tr [e^{-iH_2T}]$, for $H_1-H_2$ small. If the ensemble average of the SFF is the variance of the density fluctuations for a single Hamiltonian drawn from the ensemble, the averaged Loschmidt SFF is the covariance for two Hamiltonians drawn from a correlated ensemble. This object is a time-domain version of the parametric correlations studied in the quantum chaos and random matrix literatures. We show analytically that the averaged Loschmidt SFF is proportional to $e^{i\lambda T}T$ for a complex rate $\lambda$ with a positive imaginary part, showing in a quantitative way that the long-time details of the spectrum are exponentially more sensitive to perturbations than the short-time properties. We calculate $\lambda$ in a number of cases, including random matrix theory, theories with a single localized defect, and hydrodynamic theories.
\end{abstract}
\section{Introduction}

Given an ensemble of ``quantum chaotic'' Hamiltonians $\{H\}$, the averaged Spectral Form Factor (SFF) is defined as 
\begin{equation}
    \SFF(T) = \overline{|\tr(e^{-i H T})|^2},
\end{equation}
where the overline denotes an ensemble average. The SFF is known to exhibit a ramp-like structure at intermediate times which is characteristic of a random-matrix-like spectrum for $H$, a defining feature of quantum chaos~\cite{haake2010quantum,PhysRevLett.52.1,mehta2004random}. In this paper, we study a generalization of the SFF called the Loschmidt SFF (LSFF). The LSFF is defined in terms of two Hamiltonians $H_1$ and $H_2$ as
\begin{equation}
    \LSFF(T) = \overline{\tr(e^{i H_1 T}) \tr(e^{-i H_2 T})},
\end{equation}
where again the overline denotes an average. The goal of the paper is to motivate the study of the LSFF and to study it in a variety of representative contexts.

To explain why the LSFF is natural object to consider, let us begin with another basic feature of chaotic systems: the exponential decay of auto-correlation functions. Consider a complete set $\{A\}$ of Hermitian operators and define the infinite temperature auto-correlation function for each $A$ as
\begin{equation}
    G_{A}(T) = \frac{1}{D} \tr\left( e^{i H T} A e^{-i H T} A \right),
\end{equation}
where $H$ is the system Hamiltonian and $D$ is the Hilbert space dimension. In a chaotic system, one typically expects $G_A(T) \sim e^{- \gamma_A T} + \cdots$, at least at intermediate times. 

Rather than considering a single such $G_A$, it is often convenient to consider sum over all $A$, in which case we obtain a result proportional to the SFF,
\begin{equation}
    \sum_A G_A(t) = \SFF(T).
\end{equation}
As mentioned above, the form factor directly probes the correlations between energy levels of the Hamiltonian in an operator-independent way. Moreover, it can be argued on effective field theory grounds that exponential decay of correlations indeed implies the characteristic ``ramp'' phenomenon in the spectral form factor~\cite{winer2021hydrodynamic}.

The correlators $G_A$ can be measured in the following way. Consider two copies of the system, $S$ and $\bar{S}$, prepared in an infinite-temperature thermofield double state along with a control qubit $C$ initialized in the state $\frac{|0\rangle + |1\rangle}{\sqrt{2}}$. Using a conditional application of operator $A$, followed by an unconditional time-evolution, followed by another conditional application of $A$, the state becomes 
\begin{equation}
    \frac{1}{\sqrt{2}} \left(|0\rangle_C \otimes e^{- i H T} |\infty\rangle_{S\bar{S}} + |1\rangle_C \otimes A e^{- i H T} A |\infty\rangle_{S\bar{S}} \right).
\end{equation}
By measuring the Pauli operators $X_C$ and $Y_C$ and repeating to collect statistics, one can then estimate $G_A(T)$ via
\begin{equation}
    \langle X_C  \rangle + i \langle Y_C\rangle = G_A.
\end{equation}

From this point of view it is natural to ask what happens if the time-evolution is itself conditional. Suppose the system evolves according to $H_1$ if the control is in state $|0\rangle$ and according to $H_2$ if the control is in state $|1\rangle$. In this case, the experimental procedure now yields
\begin{equation}
    \langle X_C\rangle + i \langle Y_C\rangle = L_A(T),
\end{equation}
where $L_A(T)$ is the ``Loschmidt'' auto-correlation function,
\begin{equation}
   L_A(T) = \frac{1}{D} \tr\left( e^{i H_1 T} A e^{-i H_2 T} A \right).
\end{equation}

We refer to this as a Loschmidt auto-correlator since it is correlation function version of the traditional Loschmidt echo, which is defined by taking an initial state $|\psi\rangle$, evolving for time $T$ with Hamiltonian $H_2$, then evolving for time $T$ with Hamiltonian $-H_1$. The return amplitude is $\langle \psi | e^{i H_1 T} e^{- i H_2 T} | \psi \rangle$, and when the state $|\psi\rangle$ is the infinite temperature thermofield double state, the return amplitude is $L_{\text{Id}}(T)$. 

The Loschmidt correlator is thus a natural generalization of the return amplitude in the Loschmidt echo. Moreover, if we sum the Loschmidt correlator over all choices of $A$, we get precisely the LSFF,
\begin{equation}
    \sum_A L_A(T) = \LSFF(T).
\end{equation}
Hence, the LSFF is an object that can probe both spectral correlations and the physics of the Loschmidt echo in an operator-independent way.

As we discuss below, in addition to these elementary motivations, the LSFF appears in a variety of other contexts, including as a part of the SFF in systems with spontaneous symmetry breaking. In fact, this symmetry breaking application is how we first came to consider the LSFF. The LSFF is related to various quantities such as work statistics ~\cite{Chenu_2018,Chenu_2019}. Finally, the LSFF is a time-domain version of the long-studied phenomenon known as parametric correlations, e.g.~\cite{PhysRevLett.70.4063,Weidenmuller_2005,guhr1998random,https://doi.org/10.48550/arxiv.2205.12968}. For all these reasons, the LSFF is a natural extension of the SFF which is worthy of study in its own right.

In the remainder of the introduction, we include two subsections, one that defines the LSFF and its filtered cousins in more detail and one that discusses our motivations in more detail. The rest of the paper is organized as follows. Section \ref{sec:Loschmidt} derives a formula for the Loschmidt SFF, connecting it to the Loschmidt echo. This section also extends beyond the standard Loschmidt analysis to include more complicated connected diagrams. Section \ref{sec:hydro} reformulates these results in a hydrodynamic language, and calculates new results for the Loschmidt SFF in hydrodynamic systems with spatial extent. Finally Section \ref{sec:conclusion} contains concluding remarks.

\subsection{Random matrices and Form Factors}

Here we more thoroughly introduce the SFF and the LSFF. The energy level repulsion that is a hallmark of quantum chaos is an important prediction of random matrix theory. It is commonly diagnosed by the averaged Spectral Form Factor (SFF)~\cite{haake2010quantum,BerrySemiclassical,Sieber_2001,saad2019semiclassical,PhysRevResearch.3.023118,PhysRevResearch.3.023176}, which is defined as
\begin{equation}
    \SFF(T,f)=\overline{\tr[ f(H) e^{iHT}] \tr[ f(H) e^{-iHT}]}.
\end{equation}
Here we consider a more general definition in which we allow for a filter function $f(H)$. The filter should be some slowly varying function used to focus in on a particular energy range of interest. Useful choices include $f(H)=1, f(H)=e^{-\beta H}$ and $f(H)=\exp(-(H-E_0)^2/4\sigma^2)$.

\begin{figure}
    \centering
    \includegraphics[scale=0.75]{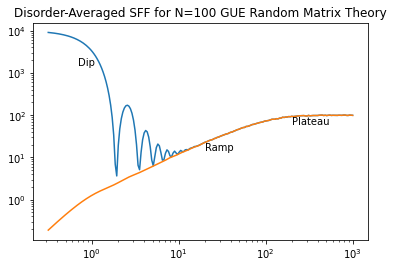}
    \caption{Full (blue) and connected (orange) SFF for GUE random matrix theory on a log-log plot. The connected SFF has the ramp-plateau structure emblematic of level repulsion and quantum chaos.}
    \label{fig:sffPic}
\end{figure}

In chaotic systems, the SFF exhibits a dip-ramp-plateau structure (figure \ref{fig:sffPic}), while in integrable systems, the ramp is not present. 
It can be shown that in systems with conserved modes, or even slow modes or nearly conserved quantities, the ramp is significantly enhanced~\cite{Friedman_2019,winer2021hydrodynamic,roy_2020,WinerGlass}.

The ensemble average, denoted by an overline, is important for rendering the averaged SFF a smooth function of time. For a single Hamiltonian, the non-averaged SFF is an erratic function of time, although an appropriate time average is typically sufficient to make it smooth. However, we are also interested in explictly disordered systems with fixed random couplings. In any case, the presence of the average means that the averaged SFF can be decomposed into the square of an average and a variance. These are often called the disconnected and connected SFF, respectively. 

Mathematically, one can write
\begin{equation}
    \SFF_{\text{conn}}=\overline{Z Z^*}-\overline{Z}\ \overline{Z^*},
\end{equation}
where 
\begin{equation}
    Z(T,f)=\sum_{n} f(E_n)\exp(-iE_nT)=\tr f(H)e^{-iHT}.
\end{equation}
$Z$ has a simple interpretation as the Fourier transform of the level density. In random matrix theory, the connected SFF has a ramp-plateau structure, with a long linear ramp terminating in a plateau (see Figure \ref{fig:sffPic}). The plateau is the variance one would get assuming random phases for the complex exponential. The ramp, where the SFF takes on a smaller value, thus represents a suppression of variation in the Fourier transform of the density. This suppression is greater at lower frequencies, owing to the long-range nature of the repulsion.

The exact Random Matrix Theory (RMT) value of the SFF in the ramp region is given by
\begin{equation}
    \SFF_{\textrm{conn}}(T,f)=\frac 1 {\bbeta \pi}\int f^2(E) dE T.
\end{equation}
Here $\bbeta$ is a quantity depending on the type of time reversal symmetry in the system. $\bbeta$ is one for real symmetric matrices with Gaussian matrices (which correspond to systems with time reversal symmetry). This ensemble is called Gaussian Orthogonal Ensemble (GOE) because the ensemble has $SO(N)$ conjugation symmetry. $\bbeta=2$ for Gaussian Hermitian matrices (which tend of correspond to systems without time reversal symmetry), which are often called Gaussian Unitary Ensemble (GUE). And $\bbeta=4$ for matrices with quaternion entries, which do emerge in physics but tend to have more obscure interpretations. These are called the Gaussian Symplectic Ensemble (GSE). Oftentimes one uses the phrase GXE to refer to the three ensembles together.

\begin{figure}
    \centering
    \includegraphics[scale=0.2]{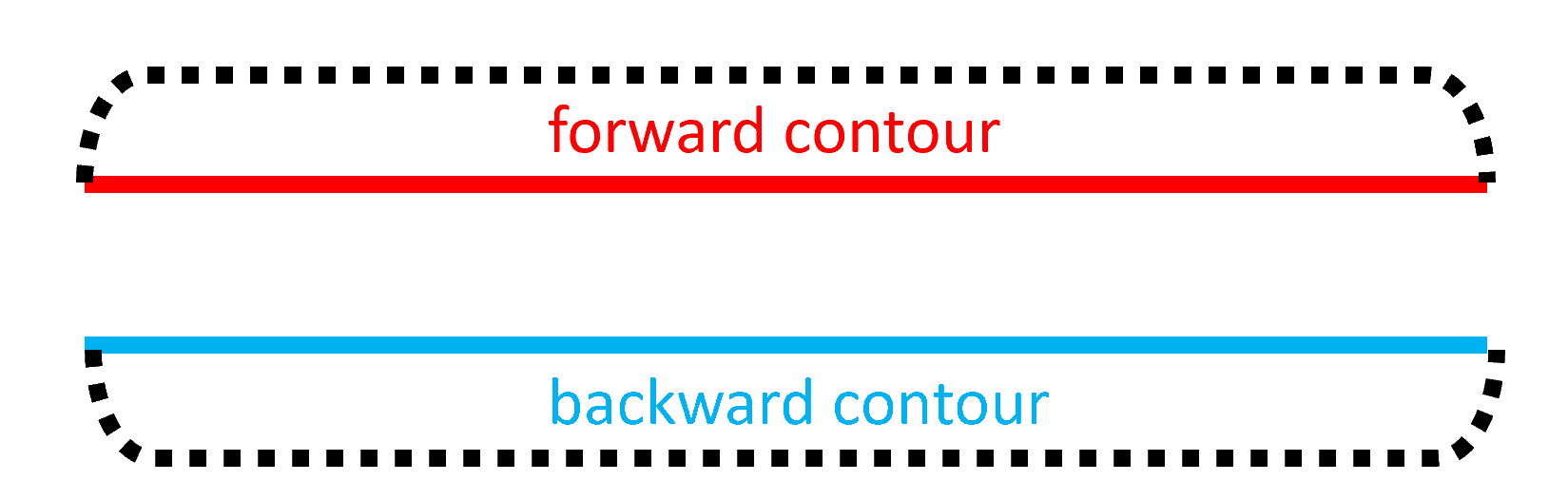}
    \caption{Two disconnected periodic contours in opposite directions. Evaluating the path integral on this contour calculates an SFF. As shown, there is no interaction between the two contours and the SFF factors, but this factorization is destroyed by the introduction of a disorder average.}
    \label{fig:sffcontour}
\end{figure}

Setting the $f$ factors equal to $1$, the SFF is a path integral on a particular time contour, depicted in Figure \ref{fig:sffcontour}.
This contour has two periodic time legs with periods $+T$ and $-T$. There is no direct interaction between them. Introducing disorder leads to an effective interaction between the two legs, allowing them to be treated with a single interacting theory \cite{saad2019semiclassical}. This theory has a $U(1)\times U(1)$ translational symmetry, one cycle for each leg of the contour. When treated semiclassically, certain saddle points spontaneously break this symmetry. The resulting Goldstone manifold has size proportional to $T$, leading to a linear ramp.

We define the (connected) Loschmidt SFF (LSFF) as
\begin{equation}
    \LSFF_{\text{conn}}=\overline{\tr[ f(H_1) e^{iH_1T}]\tr[ f(H_2) e^{-iH_2T}]}-\overline{\tr[ f(H_1) e^{iH_1T}]}\ \overline{\tr[ f(H_2) e^{-iH_2T}]}.
\end{equation}
If the connected SFF is a variance, the Loschmidt SFF is a correlation. We focus on the case where we draw two Hamiltonians $H_1,H_2$ from a joint distribution which makes them similar. For instance, they might satisfy $H_1=H+\epsilon \delta H,H_2=H-\epsilon \delta H$, for some small $\epsilon$ and $H,\delta H$ drawn from a normalized GUE ensemble. Alternatively, $H_1$ and $H_2$ might be random-field Heisenberg models or SYK clusters~\cite{sachdev_sy_1993,kitaev_syk_2015,rosenhaus_syk_2016,maldacena_syk_2016} with strongly correlated but not identical disorder. We will focus mainly on the case where $T>0$, noting that $\LSFF(-T)=\LSFF(T)^*$.

As discussed above, the name comes from an analogy with the Loschmidt echo \cite{2012,2012Loschmidt,2006Loschmidt,Goussev:2012}. The echo can be written as $|\bra \psi e^{iH_1 T}e^{-iH_2 T}\ket \psi|^2,$ and it can be interpreted as a diagnostic of the fidelity of time reversal. If one starts with a state $\ket \psi$, evolves under Hamiltonian $H_1$ for time $T$, then evolves under $-H_2\approx -H_1$ for time $T$, the Loschmidt echo diagnoses how close one comes to the original state.

\subsection{Motivations for the Loschmidt Spectral Form Factor}

There are several motivations to think about the LSFF. The most important is to answer the question of ``How different is different enough'' when it comes to fine-grained spectral statistics. This is an important question when considering ensembles of the form $H=H_0+\delta H$, where $H_0$ is some fixed large Hamiltonian and $\delta H$ is some smaller disordered perturbation. Can we think of the spectral statistics of such $H$s as independent? As we shall see, the answer for large times $T$ is yes. In physics, this ensemble has an interpretation as an ordered system with some small amount of disorder. 

In mathematics, the concept of the Dyson Process \cite{Dyson:1962brm,DysonNew,Joyner} or Matrix Brownian motion refers to starting with an initial matrix $H_1$ and adding in a GUE matrix $\delta H$ with variance proportional to some small $t$. This can be interpreted as Brownian motion in the $N\times N$ dimensional space of matrices lasting for some fictitious time $t$ with no bearing on any physical time. As the matrix evolves under this process, the eigenvalues diffuse while repelling each other. Our results show that the Fourier mode of the eigenvalue density with wavenumber $T$ decays like $\exp(-\#t|T|)$. This contrasts with pure eigenvalue diffusion which which would result in a decay like $\exp(-\#tT^2)$.

The Loschmidt SFF relates to a phenomenon called parametric correlations, e.g.~\cite{PhysRevLett.70.4063,Weidenmuller_2005,guhr1998random,https://doi.org/10.48550/arxiv.2205.12968}. Mathematicians and physicists have studied the spectral correlation functions of similar matrices since the 90s. The study of parametric correlations was typically done in the energy domain as opposed to the time domain, and was focused on systems well-described by random matrices. Our work extends it to the time domain and provides physical justification for similar results by relating them to the Loschmidt echo and to the SFF hydrodynamics \cite{winer2021hydrodynamic,Winer_2022}.

The Loschmidt SFF also emerges naturally when one is calculating full SFFs of specific systems. For systems whose Hamiltonian can be written in block diagonal form as 
\begin{equation}
    H=\begin{pmatrix}H_0-\delta H&0\\0&H_0+\delta H\end{pmatrix},
\end{equation}
the SFF naturally decomposes in the sum of the SFF of $H_0-\delta H$,$H_0+\delta H$, and twice the Loschmidt SFF between the two blocks.

Such Hamiltonians arise naturally, for instance, in the case of spontaneous symmetry breaking (SSB), where different charge sectors have very similar Hamiltonians acting on collections of 'cat states'. Indeed in \cite{Winer_2022} we performed this calculation and obtained results consistent with the more general results here. 

\section{A Simple Formula for the Loschmidt Spectral Form Factor}
\label{sec:Loschmidt}

Recall that we are interested in the quantity
\begin{equation}
    \LSFF(T,f)=\overline {\tr[f(H_1) e^{iH_1T}]\tr [f(H_2) e^{-iH_2T}]},
\end{equation}
in the case where $H_1$ and $H_2$ are two highly correlated but not identical Hamiltonians drawn from some joint distribution. We define $H_1 = H+ \epsilon \delta H$ and $H_2 = H - \epsilon \delta H$, where $\epsilon$ controls the closeness of $H_1$ and $H_2$.

Let us begin by elaborating on the similarities between this object and the traditional Loschmidt echo amplitude~\cite{2012Loschmidt,2006Loschmidt,Goussev:2012},  $\bra{\psi}e^{iH_1T}e^{-iH_2T}\ket{\psi}$. If we average the echo amplitude over $|\psi\rangle$ weighted by factors of $f(H_1)$ and $f(H_2)$, then we obtain $\tr [f(H_1)e^{iH_1T}e^{-iH_2T}f(H_2)]$. This is a single-trace version of $\LSFF(T,f)$ which can be evaluated with a Schwinger-Keldysh path integral \cite{Keldysh:1964ud,Kamenev_2009,kamenev_2011,CHOU19851,Haehl_2017}. 

\begin{figure}
    \centering
    \includegraphics[scale=0.5]{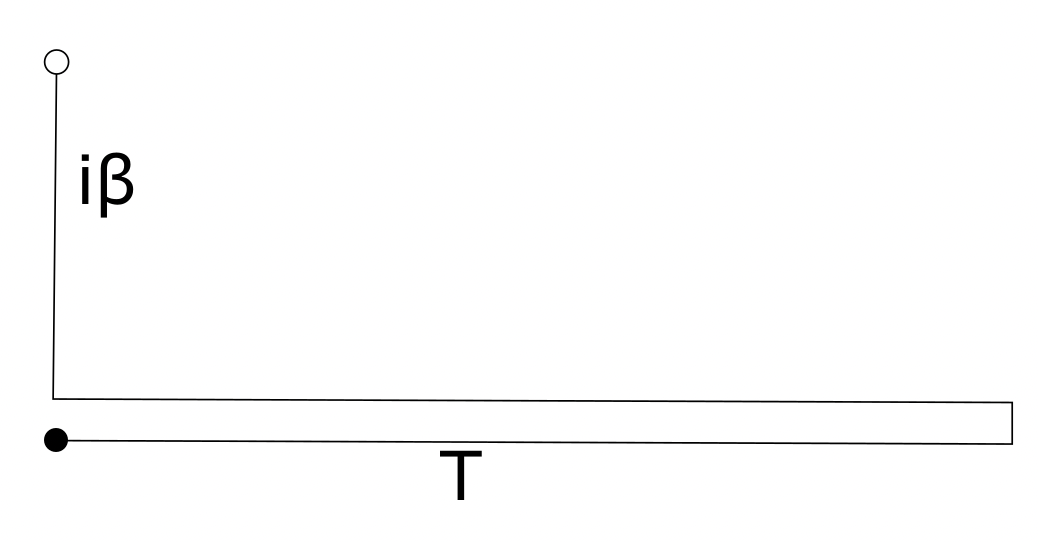}
    \caption{A Schwinger-Keldysh contour evolves forwards in time by $T$, then backwards by $T$, then by $i\beta$ in imaginary time. It is often used to calculate dynamical correlations in a thermal background. When identical sources are present on the forward and backwards legs, the partition function on this contour is equal to the traditional thermal partition function. But when the sources differ, the partition function on this contour is decreased by an amount related to the Loschmidt echo.}
    \label{fig:keldysh}
\end{figure}

The standard S-K path integral is defined on the contour depicted in Figure \ref{fig:keldysh}. The contour consists of a thermal circle to prepare a canonical ensemble, and then forwards and backward time evolution. Insertions can be placed along either or both of the real time legs of the contour. We first review this construction.

Let's assume at some distant time in the past the system is prepared in a microcanonical ensemble at energy $E$ under the Hamiltonian $H$. This can be a single state, a sampling of states from a narrow energy window, or even a thermal state. We denote the choice generically by state $|\psi\rangle$ We assume that the Hamiltonian $H$ is chaotic and obeys the eigenstate thermalization hypothesis (ETH)\cite{DeutschETH,SredETH,Rigol_2008,D_Alessio_2016}. As such, it doesn't matter very much what choice we make. Our density matrix evolves under just $H$ until time $0$ where sources $\pm \epsilon \delta H$ are turned on along the two real-time legs of the contour.

This can be evaluated to leading order in perturbation theory in $\epsilon$ in terms of a cumulant expansion~\cite{2006Loschmidt}. We want the echo amplitude,
\begin{equation}
     \overline {\mathcal P\exp(i\int_0^T \epsilon \delta H(t)-i\int_T^0 \epsilon \delta H(t))},
    \label{eq:LoschmidtFormula}
\end{equation}
where $\mathcal P$ denotes path ordering on the Schwinger-Keldysh contour, and the overline represents both a disorder average and a quantum expectation value of the operator in the interaction picture of $H$. Since equation \ref{eq:LoschmidtFormula} is the expected value of an exponential, it can be expressed as a cumulant expansion using the general identity $\overline{\mathcal P \exp(\epsilon \int O(t) dt)}=\exp(\sum_i \epsilon^i \kappa_i/i!)$, where $\kappa_i$ is the $i$th path-ordered cumulant of $\int O(t) dt$. 

The first cumulant in our expansion is just the mean,
\begin{equation}
    \overline{i\int_0^T \epsilon \delta H(t)-i\int_T^0 \epsilon \delta H(t)}=2i\epsilon \overline{\delta H}T.
\end{equation} 
The second cumulant is the variance of $\left(i\int_0^T \epsilon \delta H(t)-i\int_T^0 \epsilon \delta H(t)\right)$. Expressed in terms of 
\begin{equation}
    G(t) =\overline{\delta H(t)\delta H(0)}-\overline{\delta H}^2,
\end{equation}
where again the overline includes the quantum average, the second cumulant is
\begin{equation}
     4\epsilon^2|T| \int_{-\infty}^\infty G(t) dt
\end{equation} 
Plugging in the first two terms of the cumulant expansion gives
\begin{equation}
\begin{split}
    \bra{\psi}e^{iH_1T}e^{-iH_2T}\ket{\psi}=\exp(i\lambda T)\\
    \lambda=2\epsilon \overline {\delta H}+4i\epsilon^2\int_{0}^\infty G(t)dt+O(\epsilon^3).
\end{split}
\end{equation}
As we can see, the final answer for $\bra{\psi}e^{iH_1T}e^{-iH_2T}\ket{\psi}$ is exponential in the sum of the cumulants. This is true for exactly the same reason that the parition function is exponential in the sum of the connected diagrams.

Now we modify the construction to access the LSFF along the same lines as in \cite{winer2021hydrodynamic,saad2019semiclassical}. The filter functions can allow us to restrict the energy range and the averaging yields a non-erratic function of time and justifies coupling the two legs of the contour. The cumulant expansion on the SFF contour instead of Schwinger Keldysh gives the same cumulants, because finite-time correlations aren't sensitive to the change in boundary conditions. Plugging them into the cumulant expansion gives the novel result 
\begin{equation}
\begin{split}
    \LSFF(T,f)=|T|\int dE \frac{f^2(E)e^{i\lambda T}}{\bbeta \pi}\\
    \lambda=2\epsilon \overline {\delta H}+4i\epsilon^2\int_{0}^\infty G(t)dt+O(\epsilon^3).
\end{split}
    \label{eq:bigResult}
\end{equation}

As a comment, it will be helpful to write $\text{Im}(\lambda)$ in another form. The two-point function $4\epsilon^2\int_{-\infty}^\infty G(t)dt$ can also be written 
\begin{equation}
\begin{split}
    4\epsilon^2\int_{0}^\infty G(t)dt=4\pi \epsilon^2 \sigma^2 \rho(E)\\
    \sigma^2=|\delta H\textrm{ root-mean-square matrix element between states with energy $\sim E$}|^2.
\end{split}
\end{equation}
To give a simple example, suppose $|\psi\rangle$ is an infinite temperature state and $\delta H$ is traceless. Then the $\textrm{Im} \lambda$ term vanishes and the cumulant expansion predicts exponential decay of the echo amplitude. 

One useful analogy to have in mind is that the exponent $\lambda$ in our cumulant expansion is a type of free energy density. The Loschmidt SFF is a path integral evaluated with a nontraditional action on a nontraditional doubled spacetime of size proportional to $T$. It can be interpreted as a partition function (with imaginary temperature) of a system with Hamiltonian $H_1\otimes I-I\otimes H_2$. In this point of view, the sum of the connected diagrams is like a free energy density. To get the partition function, we multiply $\lambda$ by $T$ and exponentiate, just as we extract partition functions by doing the same to the free energy density.

\begin{figure}
    \centering
    \includegraphics[scale=0.5]{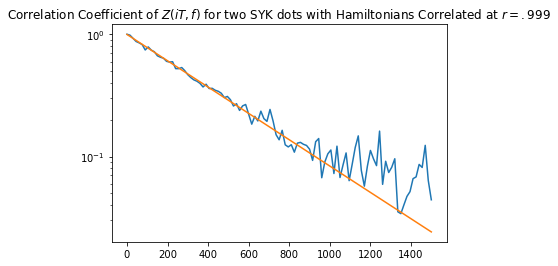}
    \caption{Loschmidt SFF divided by the SFF, numerical results (blue) vs prediction (orange). These results are for two instances of the SYK model \cite{Maldacena:2016hyu,Rosenhaus_2019,Davison_2017} with Hamiltonian $H=\sum_{ijkl}J_{ijkl}\psi_i\psi_j\psi_k\psi_l$. The two instances have $J$'s correlated with $r=.999$.}
    \label{fig:SYK}
\end{figure}

What then can we say about the more complicated connected diagrams? Whether higher cumulants can contribute meaningfully or not depends on the details of the situation in question. For instance, consider the case of a large lattice of volume $V$ with some small number of disordered defects which add some local operator $O(x)$ to the Hamiltonian. Realistically, the number of defects is proportional to $V$. The cumulants of any given $O(x)$ over time are independent of $V$, and the joint cumulants are roughly zero. So we'd expect every term in the expansion to be of order $V$, which means the Loshmidt SFF vanishes in the thermodynamic limit. To prevent this, we can either have fewer defects, or smaller defects. In the first case, we would still see every term in the expansion contributing at the same order, whereas in the second case the leading terms would dominate.
\begin{figure}
    \centering
    \includegraphics[scale=0.5]{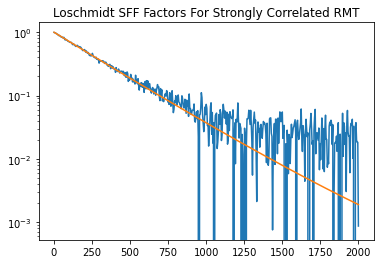}\includegraphics[scale=0.5]{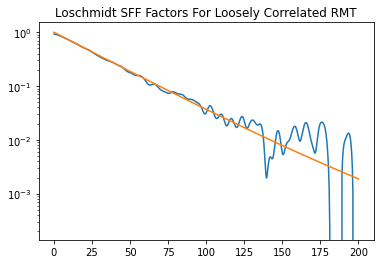}
    \caption{Loschmidt SFF divided by the SFF, numerical results (blue) vs prediction (orange). The first graph is for a sample of 1320 pairs of 1500 by 1500 GUE matrices with correlated entries at $r=.998$. The second graph is for a sample of 10830 pairs of 300 by 300 GUE matrices correlated with $r=.98$.}
    \label{fig:RMT}
\end{figure}
Figure \ref{fig:RMT} illustrates the validity of equation \eqref{eq:bigResult} for pure random matrix theory, where only the leading terms in the cumulant expansion contribute. Figure \ref{fig:SYK} shows the same for the SYK model, a fermion model with all-to-all interactions~\cite{sachdev_sy_1993,kitaev_syk_2015,rosenhaus_syk_2016,maldacena_syk_2016}.
A useful sanity check is to consider $\delta H=i[H,O]$ for some Hermitian operator $O$. Because this operator is a time derivative, we can see that $\lambda$ vanishes. This is what we expect since to leading order adding $[H,O]$ to $H$ doesn't change the spectrum, only the eigenfunctions.

\subsection{Higher Cumulants}

A useful toy model which is analytically tractable and where higher-order cumulants of $\delta H$ affect the Loschmidt SFF is obtained from an $N\times N$ GUE matrix $H$ and a projection matrix $P_k=\sum_{i=1}^k \ket{i} \bra{i}$ onto a random basis (not the energy basis) with $k \ll N$ nonzero eigenvalues. The perturbed Hamiltonians are $H_1=H_0+\epsilon P_k$ and $H_2=H_0-\epsilon P_k$. Explicitly, $H$ is a Hermitian matrix with elements chosen independently (except for $H_{0ij}=H_{0ji}^*$) such that each element is drawn from a complex distribution with complex variance $\sigma^2$. For convenience, in this section we will make the assumption that $T>0$. The $T<0$ case can be obtained from $\textrm{LSFF}(-T)=\textrm{LSFF}(T)^*$.

We will need the $n$-point functions of $\delta H=\epsilon P_k$ with a background at energy $E$. This can be evaluated by inserting $P_k=\sum_{i=1}^k \ket i\bra i $, and using $\bra {i_1} e^{-iHT}\ket {i_2}\approx \delta_{i_1,i_2}\frac{J_1(2\sqrt{N\sigma}T)}{\sqrt{N\sigma}T}$, where we make use of the fact that the states $\ket i$ are random with respect to the energy basis. As such, we have 
\begin{equation}
    G(t_1,t_2...t_n)=\epsilon^n\bra{E}P_k(t_1)P_k(t_2)...P_k(t_n)\ket{E}=k \frac {\epsilon^n}{N}e^{iEt_{1n}}\frac{ J_1(2\sqrt{N\sigma}t_{12})}{\sqrt{N\sigma}t_{12}}\frac{J_1(2\sqrt{N\sigma}t_{23})}{\sqrt{N\sigma}t_{23}}\dots \frac{ J_1(2\sqrt{N\sigma}t_{n-1n})}{\sqrt{N\sigma}t_{n-1n}},
\end{equation}
where $t_{ij}=t_i-t_j$. To leading order in $N$, these correlation functions are exactly equal to the cumulants.

\begin{figure}
    \centering
    \includegraphics{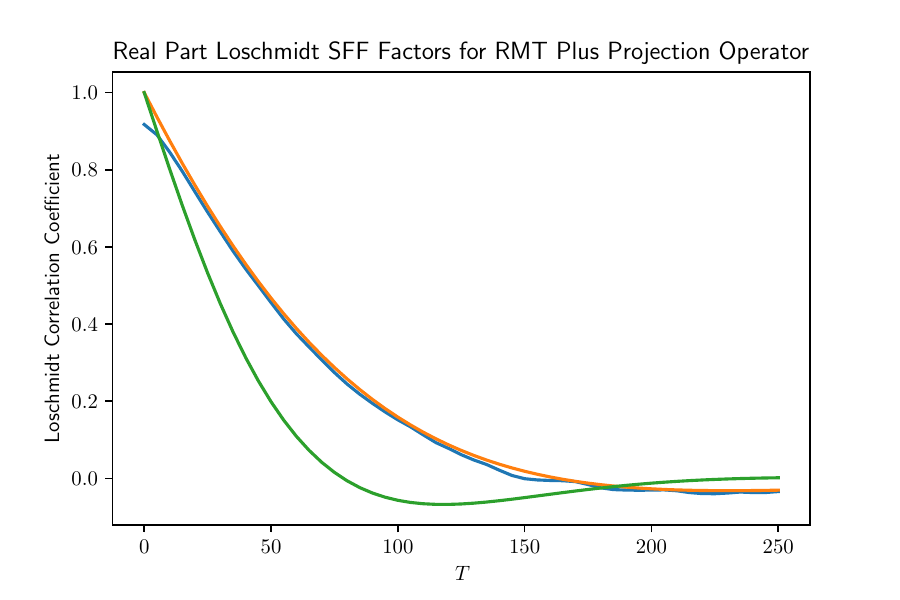}
    \includegraphics{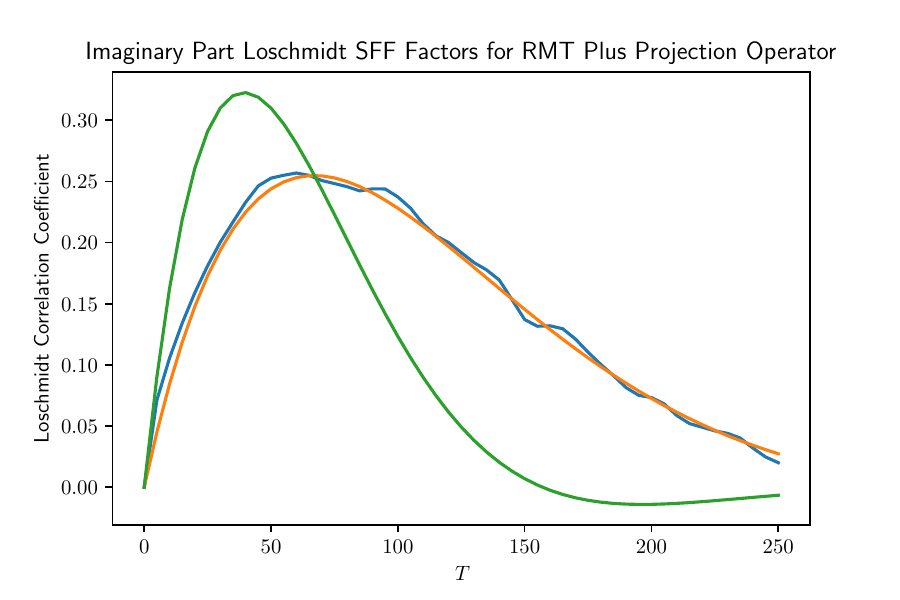}
    \caption{The numerical results (blue) line up almost perfectly with the higher-cumulant predictions of equation \eqref{eq:higherCumulantSummed} (orange), far outperforming the two-cumulant prediction (green).}
    \label{fig:cumulant}
\end{figure}

Integrating over all time-ordered configurations of $t$s we get a total contribution of
\begin{equation}
    \int d^n t\,  G(t_1,t_2...t_n)=2n k\frac {\epsilon^n}{N^n}(\pi \rho(E))^{n-1} T.
\end{equation}
The factor of $2n$ comes from the number of time-ordered ways to assign the insertions to the two contours. Once that is done, the $t_{ij}$ integrals can be done separately. Summing these contributions gives
\begin{equation}
    \begin{split}
        \lambda=2k\frac{\epsilon}{N} \frac 1 {(1-\pi i \rho(E) \epsilon/N)^2}
        \label{eq:higherCumulantSummed},
    \end{split}
\end{equation}
which can be shown to agree with equation \eqref{eq:bigResult} up to the first two orders in $\epsilon$. The accuracy of equation \eqref{eq:higherCumulantSummed} is borne out in Figure \ref{fig:cumulant}.

\subsection{The Loschmidt Spectral Correlation Function in Energy Space}

For GUE systems, the connected spectral form factor is given by
\begin{equation}
\begin{split}
    \SFF_{conn}(T,f)=\int dE f^2(E) g(T,\rho(E))\\
    g(T,\rho)=\begin{cases} 
      \frac {|T|}{2\pi} & |T|\leq 2\pi \rho \\
      \rho & |T|> 2\pi \rho
   \end{cases}.
\end{split}
\end{equation}
We can take the Fourier transform of this to get the two-point function in position space:
\begin{equation}
    \langle \rho(E)\rho(E+\Delta E)\rangle_{conn}=- \frac{\sin^2(\rho(E)\Delta E/\pi)}
{(\Delta E)^2}+\rho(E)\delta(\Delta E).
\label{eq:RMTtwopoint}
\end{equation}
Taking the same Fourier transform of equation \eqref{eq:bigResult}, we obtain the two-point function of the densities of two different Hamiltonians:
\begin{equation}
    \langle \rho_1(E)\rho_2(E+\Delta E)\rangle_{conn}=- \frac{(\Delta E+\textrm{Re }\lambda)^2-\textrm{Im }\lambda^2}
{2((\Delta E+\textrm{Re } \lambda)^2+\textrm{Im }\lambda^2)^2}.
\label{eq:Loschmidttwopoint}
\end{equation}
For large $\Delta E$, equations \eqref{eq:RMTtwopoint} and \eqref{eq:Loschmidttwopoint} agree, but the short-range behavior is entirely different, confirming that small changes to the Hamiltonian have a drastic effect at low energies (long times) but a negligible effect at high energies (short times).

\subsection{How Different Is Different Enough?}

One of the central questions we hope to answer is how different two samples need to be in order to have effectively independent spectral statistics. We now have the power to answer this question. If we have two samples with a single defect on a single site different between them, this will lead to a non-extensive decay of the Loschmidt SFF. A decay of the Loschmidt SFF independent of system size contrasts with many SFF-related quantities that grow with system size, such as the Heisenberg time which grows exponentially with system size and the Thouless time which grows (for systems with a local conserved energy) at least quadratically in system size\cite{winer2021hydrodynamic,roy_2020, Schiulaz_2019}.

If the number of defects grows with system size at all, then we find that the Loschmidt SFF decays even more quickly. This means it would be extremely difficult to observe the Loschmidt SFF directly in any large system, unless it was prepared extremely carefully. It also means that when measuring an SFF experimentally, even tiny changes (such as changing a small but extensive fraction of the couplings) to the system guarantee that the samples are effectively independent.

\section{The Loschmidt SFF for Hydrodynamic Systems}
\label{sec:hydro}

So far we primarily considered GXE Hamiltonians which serve as a toy model for chaotic quantum systems without any conserved quantities. It is also interesting to consider models with slow modes due to the presence of conserved or almost conserved quantities. In this context, one powerful technique for calculating SFFs is a formulation of hydrodynamics known as the Doubled Periodic Time (DPT) formulation \cite{winer2021hydrodynamic}. This technology is itself built around a hydrodynamic theory known as the Closed Time Path (CTP) formalism \cite{crossley2017effective,Glorioso_2017,glorioso2018lectures}. The CTP formalism is a theory of hydrodynamic slow modes on the Schwinger-Keldysh contour, and the DPT formalism transfers these results onto the SFF contour.
\begin{figure}
    \centering
    \includegraphics[scale=0.1]{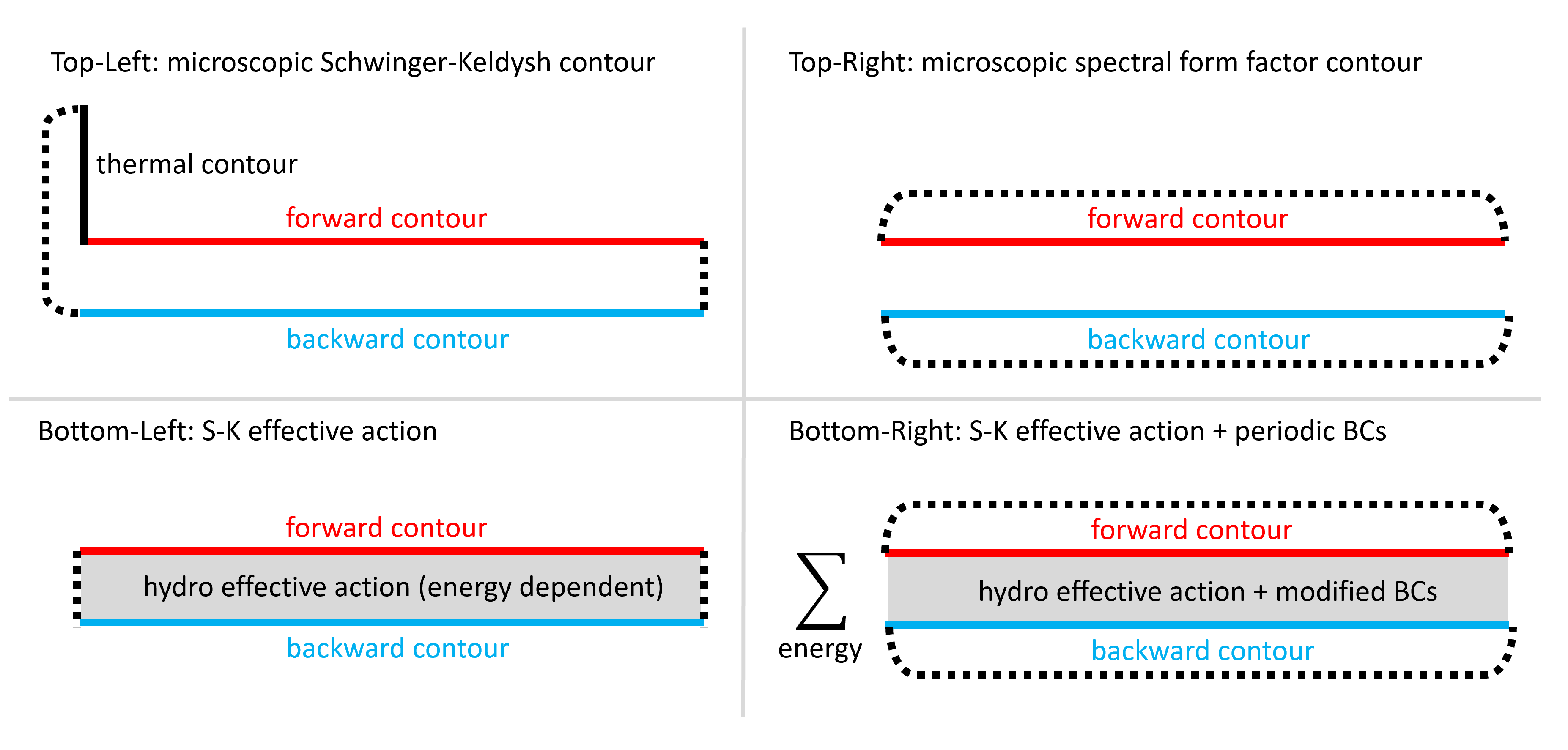}
    \caption{The idea behind the Doubled Periodic Time (DPT) formalism. Just as how microscopic fields on a Schwinger-Keldysh contour are integrated out to give an effective field theory of hydrodynamics, the microscopic degrees of freedom on an SFF contour are integrated out to give that same action, up to different boundary conditionals and exponentially small (in $T$) corrections.}
    \label{fig:BrianMasterpiece}
\end{figure}
In this section, we review the DPT formalism and then extend it to the Loschmidt SFF.

\subsection{Quick Review of CTP Formalism}
\label{subsec:CTP}
Hydrodynamics can be viewed as the program of creating effective field theories (EFTs) for systems based on the principle that long-time and long-range physics is driven primarily by conservation laws and other protected slow modes. One particular formulation is the CTP formalism explained concisely in ~\cite{glorioso2018lectures}. For more details see ~\cite{crossley2017effective,Glorioso_2017}. Additional information about fluctuating hydrodyamics can be found in~\cite{Grozdanov_2015,Kovtun_2012,Dubovsky_2012,Endlich_2013}.

The CTP formalism is the theory of the following partition function: 
\begin{equation}
    Z[A^\mu_1(t,x),A^\mu_2(t,x)]=\tr \left( e^{-\beta H} \mathcal{P} e^{i\int dt d^d x A^\mu_1j_{1\mu}} \mathcal{P} e^{-i\int dt d^d x A^\mu_2 j_{2\mu}}\right),
\end{equation}
where $\mathcal P$ is a path ordering on the Schwinger-Keldish contour. Here the $j$ operators are local conserved currents. $j_1$ and $j_2$ act on the forward and backward contours, respectively. 

For $A_1=A_2=0$, $Z$ is just the thermal partition function at inverse temperature $\beta$. Differentiating $Z$ with respect to the $A$s generates insertions of the conserved current density $j_\mu$ along either leg of the Schwinger-Keldysh contour. Thus $Z$ is a generator of all possible contour ordered correlation functions of current operators.

One always has a representation of $Z$ as a path integral,
\begin{equation}
     Z[A^\mu_1,A^\mu_2]=\int  \prod_n \mathcal D\phi^n_1\mathcal D \phi^n_2 \exp\left(i\int dt dx W_{micro}[A_{1\mu},A_{2\mu},\phi^n_1,\phi^n_2]\right),
\end{equation} 
for some collection of microscopic local fields, the $\phi^a$s. The fundamental insight of hydrodynamics is that at long times and distances, any $\phi^a$s that decay rapidly can be integrated out. What's left over is one effective $\phi$ per contour to enforce the conservation law $\partial^\mu j_{i\mu}=0$. Our partition function can thus be written 
\begin{equation}
    \begin{split}
         Z[A^\mu_1,A^\mu_2]=\int \mathcal D \phi_1\mathcal D \phi_2 \exp\left(i\int dt dx W[B_{1\mu},B_{2\mu}]\right),\\
         B_{i\mu}(t,x)=\partial_\mu \phi_i(t,x)+A_{i\mu}(t,x),
    \end{split}
    \label{eq:AbelianB}
\end{equation}
which is essentially a definition of the hydrodynamic field $\phi$ and the effective action $W$.

Insertions of the currents are obtained by differentiating $Z$ with respect to the background gauge fields $A_{i\mu}$. A single such functional derivative gives a single insertion of the current, and so one presentation of current conservation is the identity $\partial_\mu \frac{\delta Z}{\delta A_{i\mu}} = 0$. A demonstration that this enforces the conservation law is given in appendix \ref{app:CTPstuff}.

Crucially, the functional $W$ is not arbitrary. The key assumption of hydrodynamics is that $W$ is the spacetime integral of a local action. Moreover, when expressed in terms of
\begin{equation}
\begin{split}
    B_a=B_1-B_2,\\
    B_r=\frac{B_1+B_2}{2},
\end{split}
\end{equation}
there are several constraints which follow from unitarity:
\begin{itemize}
    \item $W$ terms all have at least one power of $B_a$, that is $W=0$ when $B_a=0$.
    \item Terms odd (even) in $B_a$ make a real (imaginary) contribution to the action.
    \item All imaginary contributions to the action are positive imaginary.
    \item A KMS constraint imposing fluctuation-dissipation relations.
    \item Unless the symmetry is spontaneously broken, all factors of $B_r$ have at least one time derivative.
    \item Any correlator in which the last variable has $a$ time will evaluate to 0 (known as the last time theorem or LTT).
\end{itemize}
When calculating SFFs, one typically sets the external sources $A$ to zero, so the action can be written purely in terms of the derivatives of the $\phi$s.

The $\phi$s have a physical interpretation depending on the precise symmetry in question. In the case of time translation / energy conservation, the $\phi$s are the physical time corresponding to a given fluid time (and are often denoted $\sigma$). In the case of a U(1) internal symmetry / charge conservation, they are local fluid phases. One simple quadratic action for an energy-conserving system consistent with the above conditions is
\begin{equation}
    L=\sigma_a\left(D\kappa \beta^{-1}\nabla^2\partial_t \sigma_r-\kappa \beta^{-1}\partial_t^2\sigma_r\right)
    +i\beta^{-2}\kappa(\nabla \sigma_a)^2. 
\end{equation}
Here $D$ represents the diffusion coefficient, $\kappa$ is the thermal conductivity (energy flux per temperature gradient) and $\beta$ is the equilibrium temperature.

The reader should imagine that this action is corrected by cubic and higher orders terms in the $\sigma$s and by higher derivative corrections even at the quadratic level.

\subsection{Moving on to DPT and the SFF}

As argued in \cite{winer2021hydrodynamic}, the SFF enhancement for a diffusive system can be calculated by evaluating the path integral
\begin{equation}
    \SFF = \int \mathcal{D} \sigma_1 \mathcal{D} \sigma_2 f(E_1)f(E_2)e^{i W(\sigma_1,\sigma_2)},
\end{equation}
where $\sigma_{1,2}$ represent reparameterization modes on the two legs of the contour. It is often useful to define $\sigma_a=\sigma_1-\sigma_2$ and $\rho=\frac {\partial S}{\partial(\partial_t \sigma_a)}$. Here $\rho$ is the average energy density between the two contours, and can be written $\rho=\kappa \beta^{-1}\partial_t \phi_r$. In this notation, 
\begin{equation}
    L=-\sigma_a\left(\partial_t\rho-D\nabla^2\rho\right)+i\beta^{-2}\kappa(\nabla \sigma_a)^2,
\end{equation}
and the effective action is $W = \int dt L$. Importantly, the boundary conditions are no longer those of the Schwinger-Keldysh contout but those of the SFF contour.

Since the action is entirely Gaussian, we can evaluate the path integral exactly. We first break into Fourier modes in the spatial directions. The remaining integral is
\begin{equation}
    \prod_{k}\int\mathcal D\epsilon_k\mathcal D\sigma_{ak}f(E_1)f(E_2) \exp\left(-i\int dt \sigma_{ak}
    \partial_t\rho_k+Dk^2\sigma_{ak}\rho_k-\beta^-2\kappa k^2 \sigma_{ak}^2\right) 
\end{equation}
For $k\neq 0$, breaking the path integral into time modes gives an infinite product which works out to $\frac{1}{1-e^{-Dk^2T}}$. For $k=0$, we just integrate over the full manifold of possible zero-frequency $\sigma_a$s and $\rho$s to get $\frac T{2\pi} \int f^2(E) dE$. Including other modes gives
\begin{equation}
    \SFF=\left[\prod_k \frac{1}{1-e^{-Dk^2T}}  \right]\frac T{2\pi} \int f^2(E) dE.
\end{equation}

\subsection{Coupling In Sources}

In this subsection, we will focus specifically on sources that couple to conserved currents, but the next paragraph applies to any operator. Because of the relative minus sign between the two contours, $A_r$ couples to $j_a$ and vice versa. A configuration where $A_a=0$, $A_r=A$ corresponds to unitary evolution with background potential $A$. With CTP boundary conditions, the partition function for $A_a=0$, $A_r=A$ is exactly $Z(\beta)$, irregardless of $A_r$. So when $A_a=0$, any number of derivatives with respect to $A_r$ (insertions of $j_a$) results in a correlator of zero.

With periodic boundary conditions, this is no longer entirely true. The trace can take on different values, and changing details of the unitary evolution results in a change in the SFF, albeit one which decays as $T$ grows and the effects of the periodic boundary conditions grow more mild. The intuitive explanation for this is that the SFF at times less than the Thouless time depends on non-universal properties of the Hamiltonian, and coupling in $A_r$ is effectively a change to the Hamiltonian. Thus it can affect the SFF before the Thouless time.

$A_a$ is a different story. Turning on a nonzero $A_a$ term corresponds to having a different Hamiltonian on the forward vs backwards path. This changes the partition function even in the CTP case. In the periodic-time setting, this transforms our SFF into a Loschmidt SFF, a pair of periodic contours with slightly different Hamiltonians along the two legs.

\subsection{A Perturbative Look at the Loschmidt SFF in Hydro}

In this subsection we will restrict our attention to systems with no conserved quantities besides the energy $H$, and thus only one conserved current $j_\mu$.

Assuming the perturbing $\epsilon \delta H$ has an overlap $\epsilon \delta (x-x_0)$ with the local energy density $j_0(x)$, we can model the Loschmidt SFF as
\begin{equation}
    \LSFF(T,f)=Z_{\text{DPT}}[A_{\mu r}=0, A_{\mu a} =2\epsilon\delta(x-x_0)\delta_{\mu 0}]
\end{equation}
To leading order in $\epsilon$, the `free energy' $\lambda$ is just a $-2i\epsilon j_{0r}(x_0)$ insertion, which is a typical diagonal matrix element in the energy shell, so we have
\begin{equation}
    \LSFF(T,f)=\frac T{2\pi}\int dE f^2(E)\exp\left(2i\epsilon  j_{0r}(x_0)T+O(\epsilon^2)\right).
\end{equation}

To second order in $\epsilon$, the object in the exponent is
\begin{equation}
\begin{split}
    \textrm{Re} \lambda=2\epsilon  j_{0r}(x_0)\\
    \textrm{Im} \lambda=4\epsilon^2\int_0^T dt G_{rr;DPT}(x_0,x_0,t)
\end{split}
\end{equation}
This last integrand is the correlation function of the energy density. Because the DPT correlation functions wrap around the periodic time, this is the same as the `unwound' CTP integral 
\begin{equation}
    \int_0^T dt G_{rr;DPT}(x_0,x_0,t)=\int_{-\infty}^\infty dt G_{rr;CTP}(x_0,x_0,t).
    \label{eq:gIntegral}
\end{equation}
At times below the Thouless time and assuming spatial translation symmetry, $G_{rr;CTP}(x_0,x_0,t)$ can be modeled as 
\begin{equation}
    G_{rr;CTP}(x_0-x_0,t)=\frac{\kappa}{\beta^2D}\frac{1}{\sqrt{2\pi D|T|}^d}.
\end{equation}
For $d\geq 2$, this has a UV divergence, which can be cured by imposing a UV cutoff on the extent of our operator $\delta H$. The IR divergence of $d\leq 2$ is more interesting. It is also cured by a cutoff: the system has some finite size, and the IR behavior of $G_{rr}$ depends on that system's size and shape, as well as the precise location of $x_0$ and how strong the slowest modes are there.

\subsection{Exact Evaluation for $d=1$}

As an illustration of how integral \eqref{eq:gIntegral} can depend on $x_0$, we can evaluate it exactly when we have a diffusive system in 1d with length $L$.

We first express $G_{rr;CTP}(x_0,x_0,t)$ as an infinite sum:
\begin{equation}
    G_{rr;CTP}(x_0,x_0,t)=\sum_i f_i(x_0)^2\beta^{-2}\kappa e^{-Dk_i^2 |t|},
\end{equation}
where the $f_i$s are eigenvalues of $\nabla^2$ with eigenvalues $-k_i^2$. Performing the integral gives us the sum
\begin{equation}
    \int_{-\infty}^\infty dt G_{rr;CTP}(x_0,x_0,t)=\sum_i f_i(x_0)^2\frac {2\kappa}{\beta^2D k_i^2}.
    \label{eq:inverseSum}
\end{equation}
This is just a multiple of $-(\nabla^2)^{-1}(x_0,x_0)$. We define
\begin{equation}
    C(x_1,x_2)=
    \begin{cases} 
      \frac 1L (L-x_2)x_1 & x_1\leq x_2 \\
      \frac 1L x_2 (L-x_1) & x_1\geq x_2
   \end{cases}.
\end{equation}
Then
\begin{equation}
    \nabla_1^2 C(x_1,x_2)=-\delta(x_1,x_2).
\end{equation}
So the sum in equation \eqref{eq:inverseSum} is 
\begin{equation}
    \frac{2 \kappa}{\beta^2 D}C(x_0,x_0)=\frac{2 \kappa x_0(L-x_0)}{\beta^2 DL}.
\end{equation}

\section{Conclusion and Discussion}
\label{sec:conclusion}
In this paper, we defined and studied the Loschmidt SFF. Just as the Loschmidt echo measures the similarity between $e^{iH_1T}$ and $e^{iH_2T}$ in terms of their actions on a state, the Loschmidt SFF measures their similarity in terms of spectral statistics. We found that the Loschmidt SFF decays exponentially as a function of $T$, with the same exponential rate as the echo. We studied this quantity in several situations, including an RMT model where the exponent required just a two point function, and a model requiring higher cumulants. In both cases, our analytical prediction matched up with numerical results. We also obtained analytic results about the Loschmidt SFF in theory where the slow dynamics is governed by a hydrodynamic theory of diffusion.

One natural extension of our work would be to study the Loschmidt SFF for integrable systems. In particular, does it always have the same long-time behavior as the Loschmidt echo?

Another important direction is to look for the Loschmidt analogue of the other connection between random matrix theory and quantum chaos: the eigenstate thermalization hypothesis. If an operator $O$ is written in the energy eigenbases of Hamiltonians $H_1$ and $H_2$, is there any relation between the matrix elements in the two bases? 

This work was supported by the Joint Quantum Institute (M.W.) and by the Air Force Office of Scientific Research under award number FA9550-19-1-0360 (B.S.).

Note: Just before we posted this work, we learned of an independent study of the LSFF in a holographic context that appeared a few days before~\cite{cotler_precision_2022}. 

\appendix
\section{More details on CTP}
\label{app:CTPstuff}
In this appendix we review details of the partition function
\begin{equation}
    \begin{split}
         Z[A^\mu_1,A^\mu_2]=\int \mathcal D \phi_1\mathcal D \phi_2 \exp\left(i\int dt dx W[B_{1\mu},B_{2\mu}]\right),\\
         B_{i\mu}(t,x)=\partial_\mu \phi_i(t,x)+A_{i\mu}(t,x).
    \end{split}
\end{equation}
One important property is the conservation of charge, that is the property $\partial_\mu \frac{\partial Z}{\partial A^\mu_i}=0.$
We can derive this fact as follows. Since $Z$ only depends on $A_{i\mu}$ via the combined field $B_{i\mu}$, the derivative of $Z$ with respect to $A$ reduces to a functional derivative of the action,
\begin{equation}
    \frac{\delta Z}{\delta A_{i\mu}} = \int \mathcal D \phi_1\mathcal D \phi_2 \frac{\delta }{\delta (\partial_\mu \phi_i)}\exp\left(i\int dt dx W\right).
\end{equation}
Acting with $\partial_\mu$ and suppressing integration variables, we get
\begin{equation}
    \int \partial_\mu \frac{i \delta W}{\delta (\partial_\mu \phi_i)} e^{i \int W}.
\end{equation}
Because $W$ does not depend explicitly on $\phi_i$ (only on its derivatives), this is the function integral of a functional total derivative and hence vanishes.
\bibliographystyle{ieeetr}
\bibliography{main.bib}

\begin{thebibliography}{10}

\bibitem{haake2010quantum}
F.~Haake, {\em Quantum Signatures of Chaos}.
\newblock Springer Series in Synergetics, Springer Berlin Heidelberg, 2010.

\bibitem{PhysRevLett.52.1}
O.~Bohigas, M.~J. Giannoni, and C.~Schmit, ``Characterization of chaotic
  quantum spectra and universality of level fluctuation laws,'' {\em Phys. Rev.
  Lett.}, vol.~52, pp.~1--4, Jan 1984.

\bibitem{mehta2004random}
M.~Mehta, {\em Random Matrices}.
\newblock ISSN, Elsevier Science, 2004.

\bibitem{winer2021hydrodynamic}
M.~Winer and B.~Swingle, ``Hydrodynamic theory of the connected spectral form
  factor,'' 2021.

\bibitem{Chenu_2018}
A.~Chenu, I.~L. Egusquiza, J.~Molina-Vilaplana, and A.~del Campo, ``Quantum
  work statistics, loschmidt echo and information scrambling,'' {\em Scientific
  Reports}, vol.~8, aug 2018.

\bibitem{Chenu_2019}
A.~Chenu, J.~Molina-Vilaplana, and A.~del Campo, ``Work statistics, loschmidt
  echo and information scrambling in chaotic quantum systems,'' {\em Quantum},
  vol.~3, p.~127, mar 2019.

\bibitem{PhysRevLett.70.4063}
B.~D. Simons and B.~L. Altshuler, ``Universal velocity correlations in
  disordered and chaotic systems,'' {\em Phys. Rev. Lett.}, vol.~70,
  pp.~4063--4066, Jun 1993.

\bibitem{Weidenmuller_2005}
H.~A. Weidenmüller, ``Parametric level correlations in random-matrix models,''
  {\em Journal of Physics: Condensed Matter}, vol.~17, pp.~S1881--S1887, may
  2005.

\bibitem{guhr1998random}
T.~Guhr, A.~M{\"u}ller-Groeling, and H.~A. Weidenm{\"u}ller, ``Random-matrix
  theories in quantum physics: common concepts,'' {\em Physics Reports},
  vol.~299, no.~4-6, pp.~189--425, 1998.

\bibitem{https://doi.org/10.48550/arxiv.2205.12968}
J.~Cotler and K.~Jensen, ``A precision test of averaging in ads/cft,'' 2022.

\bibitem{BerrySemiclassical}
M.~V. Berry, ``Semiclassical theory of spectral rigidity,'' {\em Proceedings of
  the Royal Society of London. Series A, Mathematical and Physical Sciences},
  vol.~400, no.~1819, pp.~229--251, 1985.

\bibitem{Sieber_2001}
M.~Sieber and K.~Richter, ``Correlations between periodic orbits and their role
  in spectral statistics,'' {\em Physica Scripta}, vol.~T90, no.~1, p.~128,
  2001.

\bibitem{saad2019semiclassical}
P.~Saad, S.~H. Shenker, and D.~Stanford, ``A semiclassical ramp in syk and in
  gravity,'' 2019.

\bibitem{PhysRevResearch.3.023118}
A.~Chan, A.~De~Luca, and J.~T. Chalker, ``Spectral lyapunov exponents in
  chaotic and localized many-body quantum systems,'' {\em Phys. Rev. Research},
  vol.~3, p.~023118, May 2021.

\bibitem{PhysRevResearch.3.023176}
S.~Moudgalya, A.~Prem, D.~A. Huse, and A.~Chan, ``Spectral statistics in
  constrained many-body quantum chaotic systems,'' {\em Phys. Rev. Research},
  vol.~3, p.~023176, Jun 2021.

\bibitem{Friedman_2019}
A.~J. Friedman, A.~Chan, A.~De~Luca, and J.~Chalker, ``Spectral statistics and
  many-body quantum chaos with conserved charge,'' {\em Physical Review
  Letters}, vol.~123, Nov 2019.

\bibitem{roy_2020}
D.~Roy and T.~Prosen, ``{Random Matrix Spectral Form Factor in Kicked
  Interacting Fermionic Chains},'' {\em Phys. Rev. E}, vol.~102, p.~060202,
  2020.

\bibitem{WinerGlass}
M.~Winer, R.~Barney, C.~L. Baldwin, V.~Galitski, and B.~Swingle, ``Spectral
  form factor of a quantum spin glass,'' 2022.

\bibitem{sachdev_sy_1993}
S.~Sachdev and J.~Ye, ``Gapless spin-fluid ground state in a random quantum
  heisenberg magnet,'' {\em Physical Review Letters}, vol.~70, p.~3339–3342,
  May 1993.

\bibitem{kitaev_syk_2015}
A.~Kitaev, ``{A simple model of quantum holography},'' in {\em KITP Progr.
  Entanglement Strongly-Correlated Quantum Matter}, 2015.

\bibitem{rosenhaus_syk_2016}
J.~Polchinski and V.~Rosenhaus, ``The spectrum in the sachdev-ye-kitaev
  model,'' {\em Journal of High Energy Physics}, vol.~2016, p.~1–25, Apr
  2016.

\bibitem{maldacena_syk_2016}
J.~Maldacena and D.~Stanford, ``Remarks on the sachdev-ye-kitaev model,'' {\em
  Physical Review D}, vol.~94, Nov 2016.

\bibitem{2012}
A.~Wisniacki, ``Loschmidt echo,'' {\em Scholarpedia}, vol.~7, no.~8, p.~11687,
  2012.

\bibitem{2012Loschmidt}
H.~Kohler and C.~Recher, ``Fidelity and level correlations in the transition
  from regularity to chaos,'' {\em EPL (Europhysics Letters)}, vol.~98,
  p.~10005, Apr 2012.

\bibitem{2006Loschmidt}
T.~Gorin, T.~Prosen, T.~H. Seligman, and M.~Žnidarič, ``Dynamics of loschmidt
  echoes and fidelity decay,'' {\em Physics Reports}, vol.~435, p.~33–156,
  Nov 2006.

\bibitem{Goussev:2012}
A.~Goussev, R.~A. Jalabert, H.~M. Pastawski, and D.~A. Wisniacki, ``{L}oschmidt
  echo,'' {\em Scholarpedia}, vol.~7, no.~8, p.~11687, 2012.
\newblock revision \#127578.

\bibitem{Dyson:1962brm}
F.~J. Dyson, ``{A Brownian-Motion Model for the Eigenvalues of a Random
  Matrix},'' {\em J. Math. Phys.}, vol.~3, no.~6, pp.~1191--1198, 1962.

\bibitem{DysonNew}
C.~H. Joyner and U.~Smilansky, ``Dyson's brownian-motion model for random
  matrix theory - revisited. with an {Appendix by Don Zagier},'' 2015.

\bibitem{Joyner}
C.~H. Joyner and U.~Smilansky, ``Dyson's brownian-motion model for random
  matrix theory - revisited. with an appendix by don zagier,'' 2015.

\bibitem{Winer_2022}
M.~Winer and B.~Swingle, ``Spontaneous symmetry breaking, spectral statistics,
  and the ramp,'' {\em Physical Review B}, vol.~105, mar 2022.

\bibitem{Keldysh:1964ud}
L.~V. Keldysh, ``{Diagram technique for nonequilibrium processes},'' {\em Zh.
  Eksp. Teor. Fiz.}, vol.~47, pp.~1515--1527, 1964.

\bibitem{Kamenev_2009}
A.~Kamenev and A.~Levchenko, ``Keldysh technique and non-linear sigma-model:
  basic principles and applications,'' {\em Advances in Physics}, vol.~58,
  p.~197–319, May 2009.

\bibitem{kamenev_2011}
A.~Kamenev, {\em Field Theory of Non-Equilibrium Systems}.
\newblock Cambridge University Press, 2011.

\bibitem{CHOU19851}
K.~chao Chou, Z.~bin Su, B.~lin Hao, and L.~Yu, ``Equilibrium and
  nonequilibrium formalisms made unified,'' {\em Physics Reports}, vol.~118,
  no.~1, pp.~1 -- 131, 1985.

\bibitem{Haehl_2017}
F.~M. Haehl, R.~Loganayagam, and M.~Rangamani, ``Schwinger-keldysh formalism.
  part i: Brst symmetries and superspace,'' {\em Journal of High Energy
  Physics}, vol.~2017, Jun 2017.

\bibitem{DeutschETH}
J.~M. Deutsch, ``Quantum statistical mechanics in a closed system,'' {\em Phys.
  Rev. A}, vol.~43, pp.~2046--2049, Feb 1991.

\bibitem{SredETH}
M.~Srednicki, ``Chaos and quantum thermalization,'' {\em Phys. Rev. E},
  vol.~50, pp.~888--901, Aug 1994.

\bibitem{Rigol_2008}
M.~Rigol, V.~Dunjko, and M.~Olshanii, ``Thermalization and its mechanism for
  generic isolated quantum systems,'' {\em Nature}, vol.~452, pp.~854--858, apr
  2008.

\bibitem{D_Alessio_2016}
L.~D{\textquotesingle}Alessio, Y.~Kafri, A.~Polkovnikov, and M.~Rigol, ``From
  quantum chaos and eigenstate thermalization to statistical mechanics and
  thermodynamics,'' {\em Advances in Physics}, vol.~65, pp.~239--362, may 2016.

\bibitem{Maldacena:2016hyu}
J.~Maldacena and D.~Stanford, ``{Remarks on the Sachdev-Ye-Kitaev model},''
  {\em Phys. Rev. D}, vol.~94, no.~10, p.~106002, 2016.

\bibitem{Rosenhaus_2019}
V.~Rosenhaus, ``An introduction to the syk model,'' {\em Journal of Physics A:
  Mathematical and Theoretical}, vol.~52, p.~323001, Jul 2019.

\bibitem{Davison_2017}
R.~A. Davison, W.~Fu, A.~Georges, Y.~Gu, K.~Jensen, and S.~Sachdev,
  ``Thermoelectric transport in disordered metals without quasiparticles: The
  sachdev-ye-kitaev models and holography,'' {\em Physical Review B}, vol.~95,
  Apr 2017.

\bibitem{Schiulaz_2019}
M.~Schiulaz, E.~J. Torres-Herrera, and L.~F. Santos, ``Thouless and relaxation
  time scales in many-body quantum systems,'' {\em Physical Review B}, vol.~99,
  May 2019.

\bibitem{crossley2017effective}
M.~Crossley, P.~Glorioso, and H.~Liu, ``Effective field theory of dissipative
  fluids,'' 2017.

\bibitem{Glorioso_2017}
P.~Glorioso, M.~Crossley, and H.~Liu, ``Effective field theory of dissipative
  fluids (ii): classical limit, dynamical kms symmetry and entropy current,''
  {\em Journal of High Energy Physics}, vol.~2017, pp.~1--44, Sep 2017.

\bibitem{glorioso2018lectures}
P.~Glorioso and H.~Liu, ``Lectures on non-equilibrium effective field theories
  and fluctuating hydrodynamics,'' 2018.

\bibitem{Grozdanov_2015}
S.~Grozdanov and J.~Polonyi, ``Viscosity and dissipative hydrodynamics from
  effective field theory,'' {\em Physical Review D}, vol.~91, p.~105031, May
  2015.

\bibitem{Kovtun_2012}
P.~Kovtun, ``Lectures on hydrodynamic fluctuations in relativistic theories,''
  {\em Journal of Physics A: Mathematical and Theoretical}, vol.~45, p.~473001,
  Nov 2012.

\bibitem{Dubovsky_2012}
S.~Dubovsky, L.~Hui, A.~Nicolis, and D.~T. Son, ``Effective field theory for
  hydrodynamics: Thermodynamics, and the derivative expansion,'' {\em Physical
  Review D}, vol.~85, p.~085029, Apr 2012.

\bibitem{Endlich_2013}
S.~Endlich, A.~Nicolis, R.~A. Porto, and J.~Wang, ``Dissipation in the
  effective field theory for hydrodynamics: First-order effects,'' {\em
  Physical Review D}, vol.~88, p.~105001, Nov 2013.

\bibitem{cotler_precision_2022}
J.~Cotler and K.~Jensen, ``A precision test of averaging in ads/cft,'' 2022.

\end{thebibliography}
\end{document}